\documentclass[a4paper,twocolumn,pra]{revtex4-1}

\usepackage{dsfont}
\usepackage{amsmath,amssymb}
\usepackage{graphicx}
\usepackage[usenames,dvipsnames]{xcolor}
\usepackage{colortbl,hhline}
\usepackage{hyperref}

\newcommand{\1}{\ensuremath{\mathds{1}}}
\newcommand{\R}{\ensuremath{\mathds{R}}}

\newcommand{\op}[1]{\ensuremath{\hat{#1}}}
\newcommand{\ii}{\ensuremath{\text{i}}}
\newcommand{\dd}{\ensuremath{\text{d}}}
\newcommand{\vect}[1]{\ensuremath{\boldsymbol{#1}}}
\newcommand{\avg}[1]{\ensuremath{{\langle#1\rangle}}}	
\newcommand{\eavg}[1]{\ensuremath{{\langle\!\langle#1\rangle\!\rangle}}}	
\newcommand{\si}[1]{\ensuremath{_{\text{#1}}}}
\newcommand{\se}[1]{\ensuremath{^{\text{#1}}}}
\newcommand{\bra}[1]{\ensuremath{{\langle#1|}}}
\newcommand{\ket}[1]{\ensuremath{{|#1\rangle}}}

\newcommand{\filleddot}[1]{\ensuremath{{\color[rgb]{#1}\bullet}}}
\newcommand{\opendot}[1]{\ensuremath{{\color[rgb]{#1}\boldsymbol{\bigcirc}}}}

\DeclareMathOperator{\Tr}{Tr}
\DeclareMathOperator{\sign}{sign}

\definecolor{cellbg}{rgb}{0.7,1,0.7}
\newcommand{\hc}{\cellcolor{cellbg}}
\definecolor{failcellbg}{rgb}{1,0.7,0.7}
\newcommand{\hf}{\cellcolor{failcellbg}}

\newcommand{\lbl}[2]{{\color{#1}\setlength{\fboxsep}{0.5mm}\fbox{#2}}}

\begin{document}

\title{Quantum State Tomography of a Single Qubit: Comparison of Methods}
\author{Roman Schmied}
\email{roman.schmied@unibas.ch}
\affiliation{Department of Physics, University of Basel, Klingelbergstrasse 82, CH--4056 Basel}
\date{\today}

\begin{abstract}
	The tomographic reconstruction of the state of a quantum-mechanical system is an essential component in the development of quantum technologies.
	We present an overview of different tomographic methods for determining the quantum-mechanical density matrix of a single qubit: (scaled) direct inversion, maximum likelihood estimation (MLE), minimum Fisher information distance, and Bayesian mean estimation (BME).
	We discuss the different prior densities in the space of density matrices, on which both MLE and BME depend, as well as ways of including experimental errors and of estimating tomography errors.
	As a measure of the accuracy of these methods we average the trace distance between a given density matrix and the tomographic density matrices it can give rise to through experimental measurements.
	We find that the BME provides the most accurate estimate of the density matrix, and suggest using either the pure-state prior, if the system is known to be in a rather pure state, or the Bures prior if any state is possible. The MLE is found to be slightly less accurate.
	We comment on the extrapolation of these results to larger systems.
\end{abstract}

\maketitle

\tableofcontents

\section{Introduction}
\label{sec:intro}

Quantum state tomography is the attempt to discover the quantum-mechanical state of a physical system, or more precisely, of a finite set of systems prepared by the same process~\cite{QuantumStateEstimation}. The experimenter acquires a set of measurements of different non-commuting observables and tries to estimate what the density matrix of the systems must have been before the measurements were made, with the goal of being able to predict the statistics of future measurements generated by the same process. In this sense, quantum state tomography characterizes a state preparation process that is assumed to be stable over time~\cite{vanEnk2013}.

In the context of the generation and characterization of non-classical states of Bose--Einstein condensates with internal degrees of freedom~\cite{Schmied2011b}, the system under study is known to be in a totally symmetric state because of its Bose symmetry. These states are usually described in terms of total-spin observables, with the effective spin length equal to half the atom number. In this restricted framework, quantum-state reconstruction is much more feasible than for general many-particle systems; for this reason, the reconstruction of spin (or pseudo-spin) density matrices is an important real-world case for quantum state tomography.
In practice, there are many different mathematical methods for determining a density matrix from a given experimental data set, yielding sometimes very different results, and it is not obvious which of these is objectively better, even when opinions and philosophical arguments are seemingly clear.

In order to see these methods more clearly and compare them, we apply them to the simplest possible quantum-mechanical problem of determining the density matrix of a two-level system (a qubit, or a spin of length 1/2), and compare the obtained results. We find that for qubits in general, Bayesian mean estimates (section~\ref{sec:BMEtomo}) are most accurate at determining a density matrix, in agreement with general statements of Refs.~\cite{BlumeKohout2010,Ng2012}. We generally consider mixed qubit states; for a review of \emph{pure} qubit state estimation, see Ref.~\cite{Bagan2005}.

The quantum-mechanical state of any two-level system can be expressed as a $2\times2$ density matrix
\begin{equation}
	\label{eq:rhor}
	\op{\rho} = \frac12 \left( \1 + x \op{\sigma}_x + y \op{\sigma}_y + z \op{\sigma}_z \right)
		= \frac12 \left( \1 + \vect{r} \cdot \vect{\op{\sigma}} \right)
\end{equation}
in terms of the Pauli matrices
\begin{align}
	\op{\sigma}_x &= \left( \begin{array}{cc} 0&1\\1&0 \end{array} \right) &
	\op{\sigma}_y &= \left( \begin{array}{cc} 0&-\ii\\\ii&0 \end{array} \right)\nonumber\\
	\op{\sigma}_z &= \left( \begin{array}{cc} 1&0\\0&-1 \end{array} \right) &
	\1 &= \left( \begin{array}{cc} 1&0\\0&1 \end{array} \right)
\end{align}
and the vectors $\vect{r}=(x,y,z)\in\R^3$ and $\vect{\op{\sigma}}=(\op{\sigma}_x,\op{\sigma}_y,\op{\sigma}_z)$. Since the eigenvalues of $\op{\rho}$ are $\lambda^{\pm} = \frac12(1\pm\sqrt{x^2+y^2+z^2})$ and must both be nonnegative, a \emph{Bloch vector} $\vect{r}$ only represents a physical (positive semi-definite) state if $\|\vect{r}\|^2 = x^2+y^2+z^2\le1$. The three-dimensional unit sphere of Bloch vectors, where every physically possible qubit density matrix can be represented as a point in space, is an appealing and convenient representation and will be used throughout this paper.

An alternative representation of a qubit density matrix is the spherical Wigner function~\cite{Dowling1994,Schmied2011b}
\begin{equation}
	W(\vartheta,\varphi) = \frac{1+\sqrt{3}(\sin\vartheta\cos\varphi,\sin\vartheta\sin\varphi,\cos\vartheta)\cdot\vect{r}}{\sqrt{8\pi}}
\end{equation}
defined as a pseudo-probability density on the \emph{surface} of the unit sphere. It encodes the direction of the vector $\vect{r}$ in the angular distribution and the length of $\vect{r}$ in the amplitude of the pseudo-probability density. This representation is convenient for longer spins, where the Bloch vector representation is unavailable.

Many characteristics of a qubit state $\op{\rho}$ can be expressed in terms of the length $r=\|\vect{r}\|$ of its Bloch vector alone, for example the \emph{quantum Fisher information}~\cite{Pezze2009} $F\si{Q}(\op{\rho})=r^2$, the \emph{purity} $\Tr(\op{\rho}^2)=(1+r^2)/2$, or the \emph{von Neumann entropy}
\begin{equation}
	\label{eq:entropy}
	S(\op{\rho})=-\frac{1+r}{2}\ln\left(\frac{1+r}{2}\right)-\frac{1-r}{2}\ln\left(\frac{1-r}{2}\right).
\end{equation}

In what follows we consider only Stern--Gerlach type measurements on a single qubit: a projective measurement along an axis $\vect{n}$ (with $\|\vect{n}\|=1$) is represented by the observable $\op{\sigma}_{\vect{n}}=\vect{n}\cdot\vect{\op{\sigma}}$, and has an expectation value
\begin{equation}
	\label{eq:expect}
	\avg{\op{\sigma}_{\vect{n}}}=\Tr(\op{\sigma}_{\vect{n}}\op{\rho})=\vect{n}\cdot\vect{r}.
\end{equation}
The probabilities for detecting the qubit in the ``up'' state $\ket{\vect{n}\!\!\uparrow}$ satisfying $\op{\sigma}_{\vect{n}}\ket{\vect{n}\!\!\uparrow}=+\ket{\vect{n}\!\!\uparrow}$, or in the ``down'' state $\ket{\vect{n}\!\!\downarrow}$ satisfying $\op{\sigma}_{\vect{n}}\ket{\vect{n}\!\!\downarrow}=-\ket{\vect{n}\!\!\downarrow}$, are
\begin{align}
	p_{\uparrow}(\vect{n}) &= \frac{1+\vect{n}\cdot\vect{r}}{2}, &
	p_{\downarrow}(\vect{n}) &= \frac{1-\vect{n}\cdot\vect{r}}{2},
\end{align}
respectively.

If we identically prepare $N_{\vect{n}}$ qubits and measure the observable $\op{\sigma}_{\vect{n}}$ on each one, we will find $N_{\vect{n}\uparrow}$ qubits in the $\ket{\vect{n}\!\!\uparrow}$ state and $N_{\vect{n}\downarrow}$ qubits in the $\ket{\vect{n}\!\!\downarrow}$ state, giving an estimate of the expectation value (sample mean)
\begin{equation}
	\label{eq:measureavg}
	\eavg{\op{\sigma}_{\vect{n}}}
	= \frac{N_{\vect{n}\uparrow}-N_{\vect{n}\downarrow}}{N_{\vect{n}\uparrow}+N_{\vect{n}\downarrow}}
	= \frac{N_{\vect{n}\uparrow}-N_{\vect{n}\downarrow}}{N_{\vect{n}}}.
\end{equation}
A statistical estimate of the error of this expectation value is given by the width of a binomial distribution with the same expectation value,
\begin{equation}
	\label{eq:measureerr}
	\Delta\eavg{\op{\sigma}_{\vect{n}}}
	= \frac{2\sqrt{N_{\vect{n}\uparrow}N_{\vect{n}\downarrow}}}{N_{\vect{n}}^{3/2}}.
\end{equation}
This error measure will be justified below through Eq.~\eqref{eq:loglikapprox}.

In the absence of prior knowledge about the experimental system's state, the precision of the tomographic methods of section~\ref{sec:methods} is highest if the measurement axes are arranged uniformly on the sphere. As detailed in appendix~\ref{app:uniform}, for a spin-1/2 system any angular distribution is considered uniform if its quadrupolar component vanishes. Examples of uniform sampling strategies according to this criterion are equal sampling along the three Cartesian axes, the four axes through the vertices of a tetrahedron~\cite{Rehacek2004}, or a completely uniform distribution of measurement axes over the entire sphere. In what follows, we assume that the experimenter performs the same number of single-qubit measurements along each of the three Cartesian axes $\vect{n}=\vect{e}_x$, $\vect{e}_y$, $\vect{e}_z$, which is the simplest complete and uniform measurement strategy~\cite{Scott2006}. Using different axes or more than three axes generally makes all of the following tomography schemes more complicated, and if the measurements are not uniformly distributed (for example by making more measurements along $\vect{e}_z$ than along $\vect{e}_x$ or $\vect{e}_y$) the tomographic result will generally be less precise or even biased. However, if each measurement axis is adaptively chosen depending on the previous measurement results, efficiency can be improved over the Cartesian axes~\cite{Embacher2004,Mahler2011,Kravtsov2013}. Also, multi-qubit joint measurements may yield information faster than sequential single-qubit measurements~\cite{Massar1995,Keyl2001}. Such adaptations are not considered in the present work.

If our qubits are all in the state of Eq.~\eqref{eq:rhor} and we perform $N_x$ measurements along the $x$-axis, $N_y$ along the $y$-axis, and $N_z$ along the $z$-axis, the probability of getting a certain set of results is
\begin{multline}
	\label{eq:prob}
	\mathcal{P}(N_{x\uparrow},N_{x\downarrow},N_{y\uparrow},N_{y\downarrow},N_{z\uparrow},N_{z\downarrow} | \op{\rho}) =\\
	\binom{N_x}{N_{x\uparrow}} \left(\frac{1+x}{2}\right)^{N_{x\uparrow}} \left(\frac{1-x}{2}\right)^{N_{x\downarrow}}\\
	\times \binom{N_y}{N_{y\uparrow}} \left(\frac{1+y}{2}\right)^{N_{y\uparrow}} \left(\frac{1-y}{2}\right)^{N_{y\downarrow}}\\
	\times \binom{N_z}{N_{z\uparrow}} \left(\frac{1+z}{2}\right)^{N_{z\uparrow}} \left(\frac{1-z}{2}\right)^{N_{z\downarrow}},
\end{multline}
where $N_x=N_{x\uparrow}+N_{x\downarrow}$ etc., and we will assume $N_x=N_y=N_z$ below. In such a setup, the problem of quantum state tomography is to invert Eq.~\eqref{eq:prob}: given a set of experimental results, what can we say about the qubits' density matrix that has given rise to these results? In what follows, we first present several tomographic methods and apply them to a single qubit (section~\ref{sec:methods}), make some comments about experimental and tomographic errors (section~\ref{sec:errors}), and then compare the accuracies of the different methods (section~\ref{sec:compare}).

\section{Tomographic methods}
\label{sec:methods}

\subsection{Direct inversion tomography}
\label{sec:directtomo}

The simplest tomographic method, called a \emph{direct inversion}, assumes that the sample mean $\eavg{\op{\sigma}_{\vect{n}}}$ is a good and unbiased estimate of the population mean $\avg{\op{\sigma}_{\vect{n}}}$~\cite{Newton1968}. Combining Eqs.~\eqref{eq:expect} and~\eqref{eq:measureavg} along the three Cartesian axes fully defines an estimate of the qubits' Bloch vector,
\begin{equation}
	\label{eq:rd}
	\vect{r}\si{d} = \left(
	\frac{N_{x\uparrow}-N_{x\downarrow}}{N_{x\uparrow}+N_{x\downarrow}},
	\frac{N_{y\uparrow}-N_{y\downarrow}}{N_{y\uparrow}+N_{y\downarrow}},
	\frac{N_{z\uparrow}-N_{z\downarrow}}{N_{z\uparrow}+N_{z\downarrow}}
	\right).
\end{equation}
We note that $\vect{r}\si{d}$ is the global maximum of Eq.~\eqref{eq:prob}, which is a definition of $\vect{r}\si{d}$ that is readily extensible to different measurement schemes.

\begin{figure*}
	\begin{center}
		\includegraphics[width=0.98\textwidth]{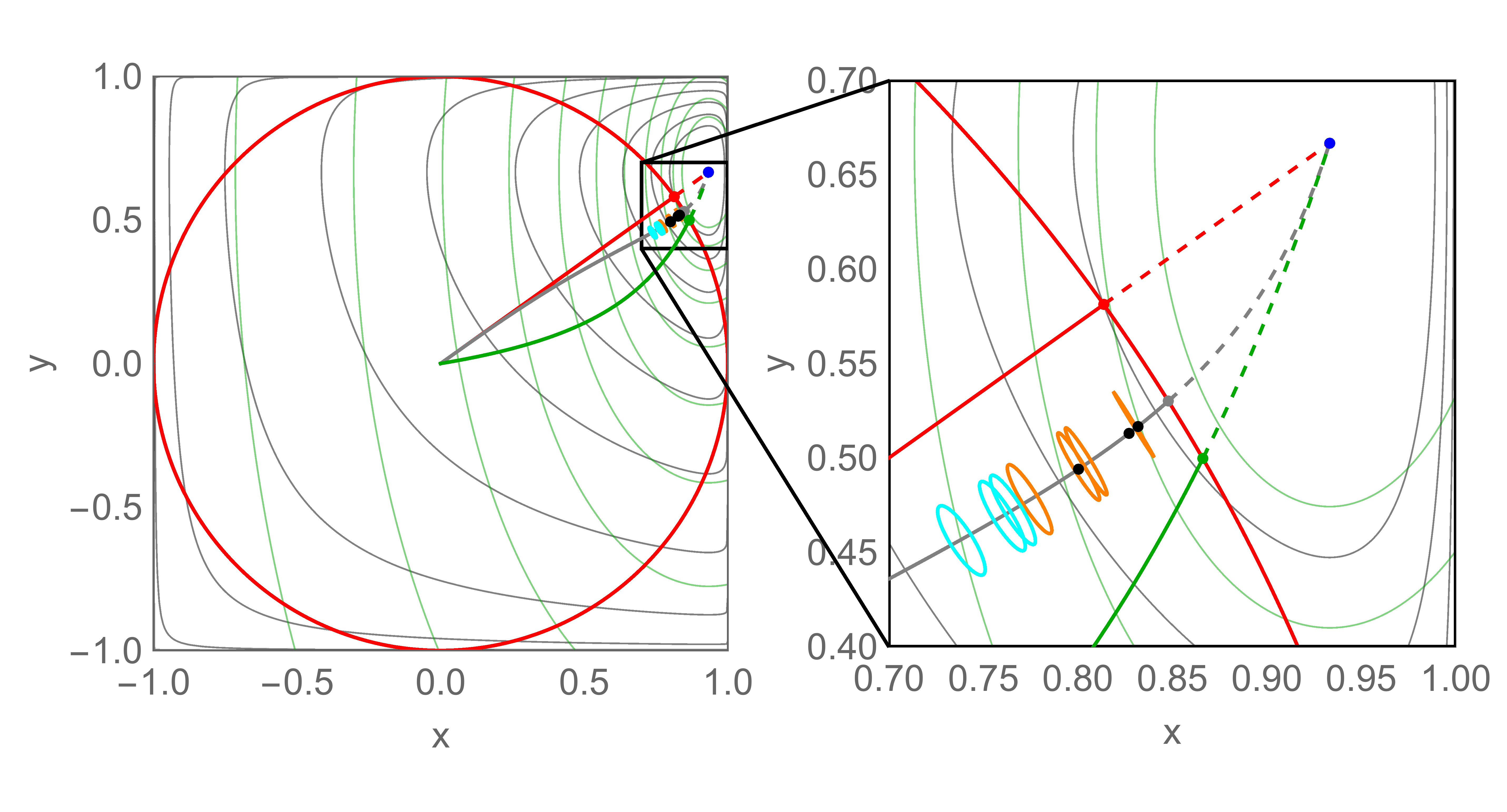}
		\caption{Example of a qubit tomography, assuming that 30 ideal measurements ($\eta=1$, see section~\ref{sec:smoothing})
		along each Cartesian quantization axes have resulted in $(N_{x\uparrow},N_{x\downarrow},N_{y\uparrow},N_{y\downarrow},N_{z\uparrow},N_{z\downarrow})=(29,1,25,5,15,15)$.
		All quantities are restricted to the $z=0$ plane.
		The \lbl{blue}{blue dot} shows the zero of the Kullback--Leibler divergence (log-likelihood) at $\vect{r}\si{d}=(\frac{14}{15},\frac{2}{3},0)$ from Eq.~\eqref{eq:rd}, which is outside of the physically allowed region $\|\vect{r}\|\le1$ indicated by the \lbl{red}{red circle}.
		The \lbl{red}{straight red line} (sections~\ref{sec:pdistance} and~\ref{sec:Schatten}) connects $\vect{r}\si{d}$ with the totally mixed state $\vect{r}=0$; the \lbl{red}{red dot} shows the result of linear scaling $\vect{r}\si{sd}=(0.814,0.581,0)$ [Eq.~\eqref{eq:fidelitymax}].
		The \lbl{Gray}{gray contours} of the Kullback--Leibler divergence are at $D\si{KL}(\vect{r}\si{d}|\vect{r})=10^{n/4}$ for $n=1\ldots10$ (outward from the blue dot).
		The \lbl{Gray}{gray line} (section~\ref{sec:likelihood}) traces the constrained likelihood maximum, Eq.~\eqref{eq:MLtrace}, as a function of $\|\vect{r}\|$; the likelihood maximum must be on this line if the prior depends only on $\|\vect{r}\|$ (Haar measure), such as the \lbl{Gray}{gray dot} showing the maximum of the likelihood at $\vect{r}\si{MLE}^{1< k\le2}=\vect{r}\si{MLE}\se{Ch}=(0.848,0.530,0)$, or the \lbl{black}{black dots} showing the maximum of the likelihood with entropy weight~\eqref{eq:entropyprior} for the Hilbert--Schmidt prior ($k=2$) at $(0.800,0.494,0)$, the Bures prior ($k=\frac32$) at $(0.827,0.513,0)$, and the Chernoff-information prior at $(0.832,0.517,0)$.
		The \lbl{Green}{green contours} of the Fisher information distance (section~\ref{sec:chi2tomo}) are at $D_{\mathcal{F}}(\vect{r}-\vect{r}\si{d})=10^{n/4}$ for $n=1\ldots11$; the \lbl{Green}{green dot} shows the point with minimum Fisher information distance at $\vect{r}\si{Fi}=(0.866,0.500,0)$, located on the \lbl{Green}{green line} of points tracing the minimum of the Fisher information distance as a function of $\|\vect{r}\|$.
		The \lbl{orange}{orange ellipses} show the Bayesian means and variances of $\vect{r}$ weighted by the likelihood (section~\ref{sec:BMEtomo}): from left to right, they use radial priors with $k=2$ (Hilbert--Schmidt measure), $k=\frac32$ (Bures measure), the Chernoff-information measure~\eqref{eq:Chernoff}, and $k=1$ (pure states only);
		the \lbl{cyan}{cyan ellipses} show the same with entropy weight~\eqref{eq:entropyprior}: from left to right, they use radial priors with $k=2$ (Hilbert--Schmidt measure), $k=\frac32$ (Bures measure), and the Chernoff-information measure.}
		\label{fig:loglik}
	\end{center}
\end{figure*}

\begin{figure*}
	\begin{center}
		\includegraphics[width=0.98\textwidth]{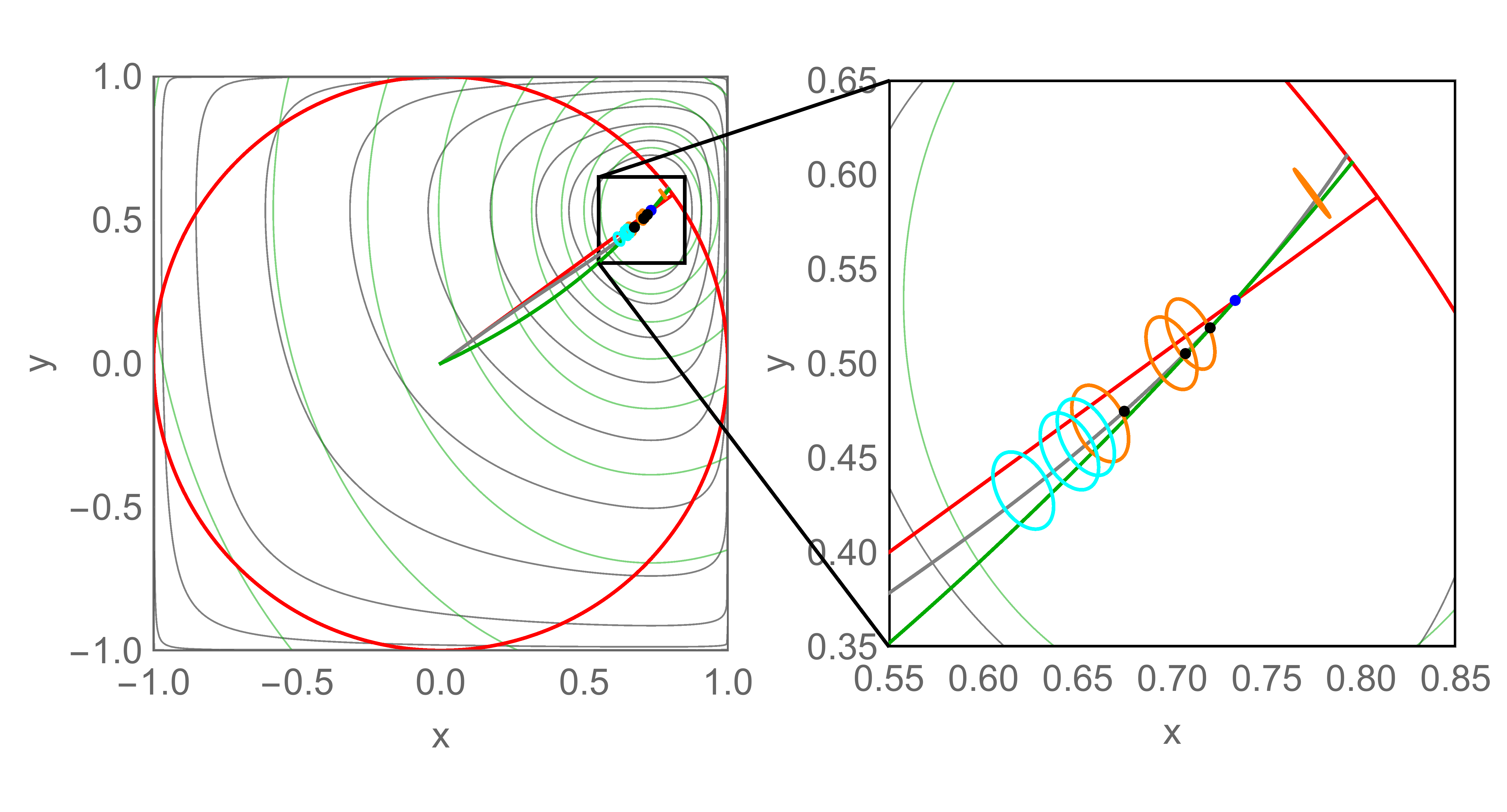}
		\caption{Same as figure~\ref{fig:loglik} but for $(N_{x\uparrow},N_{x\downarrow},N_{y\uparrow},N_{y\downarrow},N_{z\uparrow},N_{z\downarrow})=(26,4,23,7,15,15)$. The global likelihood maximum $\vect{r}\si{d}=(0.733,0.533,0)$ (blue dot) now lies inside the unit sphere and coincides with the red, gray, and green dots of figure~\ref{fig:loglik}.}
		\label{fig:loglik2}
	\end{center}
\end{figure*}

In figures~\ref{fig:loglik} and~\ref{fig:loglik2} this direct inversion Bloch vector is shown as a blue dot for two sets of experimental results, both found by performing 30 Stern--Gerlach measurements along each Cartesian axis. While in figure~\ref{fig:loglik2} the Bloch vector is physically valid since $\|\vect{r}\si{d}\|\le1$, the Bloch vector in figure~\ref{fig:loglik} is invalid and points out a fundamental problem with the direct inversion method. Eq.~\eqref{eq:rd} can be seen as three individual  parameter estimations for the three Cartesian components of the Bloch vector, and even though each parameter estimate is unconstrained on its own, the three estimates must satisfy the joint constraint $x\si{d}^2+y\si{d}^2+z\si{d}^2\le1$.
For any given state $\vect{r}$ of the qubit and for any number of measurements $(N_x,N_y,N_z)$, there is a finite probability that direct inversion tomography will find a physically invalid Bloch vector that violates this joint constraint. For example, for the completely mixed state $\op{\rho}=\frac12\1$ with $\vect{r}=(0,0,0)$, measuring $N_x=N_y=N_z=30$ times along each Cartesian direction, the probability of finding an unphysical $\vect{r}\si{d}$ is only $3\times10^{-7}$; but if we do the same measurements on the pure state $\op{\rho}=\ket{z\!\!\uparrow}\bra{z\!\!\uparrow\!\!}$ with $\vect{r}=(0,0,1)$, the chance of finding an unphysical $\vect{r}\si{d}$ is 98\%. For higher-dimensional quantum systems, this problem becomes even more severe (see section~\ref{sec:largersystems}). It has been argued recently~\cite{Schwemmer2015} that the direct inversion method provides more accurate results because it is less biased than other methods (see table~\ref{tb:ellipses} for an example of such biases); but we side with Ref.~\cite{Shang2014} in preferring physically valid density matrices despite their bias, and do not report direct inversion results in our comparison of methods. Many interesting quantities derived from the density matrix, particularly ones that go beyond linear operator expectation values and involve the entire density matrix, cannot be defined properly for density matrices that are not positive semi-definite.

Nevertheless, $\vect{r}\si{d}$ is an important starting point for many other tomographic techniques.
In what follows, we broadly distinguish between tomographic methods that minimize some distance between $\vect{r}\si{d}$ and the space of physically valid tomographic Bloch vectors (sections~\ref{sec:distancemin} and~\ref{sec:MLprior}), and methods not based on $\vect{r}\si{d}$ at all (section~\ref{sec:BMEtomo}).

\subsection{Distance minimization to $\vect{r}\si{d}$}
\label{sec:distancemin}

In order to find a valid tomographic density matrix even if $\|\vect{r}\si{d}\|>1$, we search for a modified Bloch vector $\vect{r}\si{tomo}$ that (i) is physically valid, $\|\vect{r}\si{tomo}\|\le1$, and that (ii) lies \emph{closest} to $\vect{r}\si{d}$ in terms of a distance to be defined. In figure~\ref{fig:loglik} the three dashed lines emanating from the blue dot indicate the locations of the points that minimize three types of distances to $\vect{r}\si{d}$ on concentric spherical shells around the origin ($\vect{r}=0$); their intersections with the unit sphere surface (red circle), among others, provide useable tomographic Bloch vectors, and are discussed in detail below.

\subsubsection{Minimum $p$-distance of the Bloch vectors}
\label{sec:pdistance}

The simplest family of distances between two \emph{Bloch vectors} are the $p$-distances $\|\vect{r}-\vect{r}'\|_p=(|x-x'|^p+|y-y'|^p+|z-z'|^p)^{1/p}$ for $p\ge1$. Even though the direct inversion Bloch vector $\vect{r}\si{d}$ can be located anywhere in the unit cube, the space of physically valid Bloch vectors has an intrinsic spherical symmetry around the fully mixed state $\vect{r}=0$, which suggests that only the Euclidean distance $p=2$ is to be used. In this case, the \emph{scaled direct inversion} Bloch vector minimizing the Euclidean distance to $\vect{r}\si{d}$ over the space of physically valid Bloch vectors is
\begin{equation}
	\label{eq:fidelitymax}
	\vect{r}\si{sd} = \begin{cases}
		\vect{r}\si{d} & \text{if $\|\vect{r}\si{d}\|\le1$,}\\
		\vect{r}\si{d}/\|\vect{r}\si{d}\| & \text{if $\|\vect{r}\si{d}\|>1$.}
	\end{cases}
\end{equation}
Radial scaling is shown in figures~\ref{fig:loglik} and~\ref{fig:loglik2} as a red line, with $\vect{r}\si{sd}$ indicated as a red dot.

\subsubsection{Minimum Schatten $p$-distance of the density matrices}
\label{sec:Schatten}

The simplest family of distances between two \emph{density matrices} are the Schatten $p$-distances $\|\op{\rho}-\op{\rho}'\|_p$. They include the trace distance ($p=1$) and the Frobenius or Hilbert--Schmidt distance ($p=2$). The Schatten $p$-distance between two qubit density matrices $\op{\rho}=\frac12(\1+\vect{r}\cdot\vect{\op{\sigma}})$ and $\op{\rho}'=\frac12(\1+\vect{r}'\cdot\vect{\op{\sigma}})$ is $\|\op{\rho}-\op{\rho}'\|_p = 2^{1/p}\frac12\|\vect{r}-\vect{r}'\|$, proportional to the Euclidean distance between their Bloch vectors. The minimum of any Schatten $p$-distance between the direct inversion tomography and the space of physically valid density matrices is therefore given by Eq.~\eqref{eq:fidelitymax}.

\subsubsection{Maximum fidelity}
\label{sec:fidelity}

The fidelity $F(\op{\rho},\op{\rho}')=\Tr(\sqrt{\sqrt{\op{\rho}}\cdot\op{\rho}'\cdot\sqrt{\op{\rho}}})$ is a frequently used measure of the overlap between two qubit density matrices~\cite{Schwemmer2015}. Since it does not break the spherical symmetry of the space of Bloch vectors, maximizing the fidelity between two density matrices necessarily reduces to the purely radial scaling of Eq.~\eqref{eq:fidelitymax}.

\subsubsection{Kullback--Leibler divergence and the maximum likelihood estimate}
\label{sec:likelihood}

Bayes' theorem states that if we are given a set of experimental measurements $(N_{x\uparrow}, N_{x\downarrow}, N_{y\uparrow}, N_{y\downarrow}, N_{z\uparrow}, N_{z\downarrow})$, the likelihood that a certain density matrix $\op{\rho}=\frac12(\1+\vect{r}\cdot\vect{\op{\sigma}})$ was at the source of these data is
\begin{multline}
	\label{eq:lik}
	\mathcal{L}(\op{\rho} | N_{x\uparrow},N_{x\downarrow},N_{y\uparrow},N_{y\downarrow},N_{z\uparrow},N_{z\downarrow})\\
	\propto \mathcal{C}(\op{\rho})\times \mathcal{P}(N_{x\uparrow},N_{x\downarrow},N_{y\uparrow},N_{y\downarrow},N_{z\uparrow},N_{z\downarrow} | \op{\rho}),
\end{multline}
where $\mathcal{C}(\op{\rho})$ is a prior density in the space of density matrices, vanishing whenever $\|\vect{r}\|>1$. Choosing a prior density can be a matter of taste or actual prior knowledge; however, in almost all cases the prior density will depend only on $\|\vect{r}\|$ but not on the direction of $\vect{r}$ (i.e., it is a Haar measure with respect to the spherical symmetry group).

In this section we only use the Hilbert--Schmidt measure
\begin{equation}
	\label{eq:HSprior}
	\mathcal{C}\si{HS}(\vect{r})=\begin{cases}
		\text{const.} & \text{if $\|\vect{r}\|\le1$,}\\
		0 & \text{if $\|\vect{r}\|>1$}
	\end{cases}
\end{equation}
as a prior density, which is uniform when viewed as the density of Bloch vectors within the unit sphere (but non-uniform when viewed in any other parametrization). While this is a simple and very common (often tacit) choice, it is not the most natural prior density; in section~\ref{sec:MLprior} we discuss different prior densities and their application.

A popular tomography method is to search for the maximum of the likelihood~\eqref{eq:lik} with $\mathcal{C}\si{HS}(\vect{r})$~\cite{HradilQSE3}. Since the global maximum of the probability $\mathcal{P}$, Eq.~\eqref{eq:prob}, is at $\vect{r}\si{d}$, we see that whenever $\|\vect{r}\si{d}\|\le1$ the maximum-likelihood estimate (MLE) of the Bloch vector is simply $\vect{r}\si{MLE}=\vect{r}\si{d}$. If $\|\vect{r}\si{d}\|>1$, on the other hand, we define the scaled log-likelihood, relative entropy, or Kullback--Leibler divergence~\cite{Kullback1951,HradilQSE3}
\begin{multline}
	\label{eq:loglik}
	D\si{KL}(\vect{r}\si{d}|\vect{r})=\ln\left[\frac{\mathcal{P}(\vect{r}\si{d})}{\mathcal{P}(\vect{r})}\right]\\
	= N_{x\uparrow}\ln\left(\frac{1+x\si{d}}{1+x}\right) + N_{x\downarrow}\ln\left(\frac{1-x\si{d}}{1-x}\right)\\
	+ N_{y\uparrow}\ln\left(\frac{1+y\si{d}}{1+y}\right) + N_{y\downarrow}\ln\left(\frac{1-y\si{d}}{1-y}\right)\\
	+ N_{z\uparrow}\ln\left(\frac{1+z\si{d}}{1+z}\right) + N_{z\downarrow}\ln\left(\frac{1-z\si{d}}{1-z}\right)
\end{multline}
and minimize this distance over the space of physically valid density matrices $\|\vect{r}\|\le1$~\footnote{Even though the Kullback--Leibler divergence is not a distance because it is not symmetric in its arguments, we can still minimize it with respect to one of its arguments since it is a premetric.}. Especially for large numbers of experimental data, the log-likelihood is easier to calculate in practice than the likelihood, as its dynamic range is much smaller; since the logarithm is monotonic, maximizing $\mathcal{P}$ is equivalent to minimizing $D\si{KL}$. In figures~\ref{fig:loglik} and~\ref{fig:loglik2}, the gray contours show the Kullback--Leibler divergence, and the gray dot in figure~\ref{fig:loglik} gives the likelihood maximum within the unit sphere. 
The gray line emanating from the blue dot is found by maximizing Eq.~\eqref{eq:lik}, or minimizing Eq.~\eqref{eq:loglik}, for constant $\|\vect{r}\|$, assuming that the prior depends only on $\|\vect{r}\|$: we find that these extrema are located at
\begin{equation}
	\label{eq:MLtrace}
	\vect{r}\si{MLE}(\alpha)=\left[J_{\alpha/N_x}(x\si{d}),J_{\alpha/N_y}(y\si{d}),J_{\alpha/N_z}(z\si{d})\right]
\end{equation}
with the analytic function $J_u(t)$ defined piecewise,
\begin{widetext}
\begin{equation}
	J_u(t) = \begin{cases}
		\sign(t) & \text{for $u\to-\infty$}\\
		2\sqrt{\frac{u+1}{3u}}\sign(t)\cos\left[\frac13\cos^{-1}\left(\frac32 |t| \sqrt{\frac{3u}{(u+1)^3}}\right)\right] & \text{if $u<-1$ (branch cut at $t=0$)}\\
		\sign(t)|t|^{1/3} & \text{if $u=-1$}\\
		2\sqrt{\frac{u+1}{-3u}}\sinh\left[\frac13\sinh^{-1}\left(\frac32 t \sqrt{\frac{-3u}{(u+1)^3}}\right)\right] & \text{if $-1<u<0$}\\
		t & \text{if $u=0$}\\
		2\sqrt{\frac{u+1}{3u}}\sin\left[\frac13\sin^{-1}\left(\frac32 t \sqrt{\frac{3u}{(u+1)^3}}\right)\right] & \text{if $u>0$}\\
		0 & \text{for $u\to+\infty$.}
	\end{cases}
\end{equation}
\end{widetext}
The gray line given by Eq.~\eqref{eq:MLtrace} has the following properties as a function of the Lagrange multiplier $\alpha$:
\begin{itemize}
	\item $\vect{r}\si{MLE}(\alpha)$ maximizes $\mathcal{P}(N_{x\uparrow},N_{x\downarrow},N_{y\uparrow},N_{y\downarrow},N_{z\uparrow},N_{z\downarrow} | \op{\rho})$ and minimizes $D\si{KL}(\vect{r}\si{d}|\vect{r})$ under the constraint that $\|\vect{r}\|=\|\vect{r}\si{MLE}(\alpha)\|$,
	\item $\|\vect{r}\si{MLE}(\alpha)\|$ decreases monotonically with $\alpha\in\R$,
	\item $\lim_{\alpha\to-\infty}\vect{r}\si{MLE}(\alpha)=\left(\sign[x\si{d}],\sign[y\si{d}],\sign[z\si{d}]\right)$,
	\item $\vect{r}\si{MLE}(0)=\vect{r}\si{d}$,
	\item $\lim_{\alpha\to\infty}\vect{r}\si{MLE}(\alpha)=0$.
\end{itemize}
Since $J_u(t)$ has a branch cut discontinuity at $t=0$ for $u<-1$, we must be careful when evaluating Eq.~\eqref{eq:MLtrace} if any of the $(x\si{d},y\si{d},z\si{d})$ are zero (see below).

Thus the maximum likelihood method for Cartesian-axes qubit tomography is simpler than the $\hat{R}\cdot\hat{\rho}\cdot\hat{R}$ iteration used for larger systems~\cite{HradilQSE3}, and consists of the following steps:
\begin{enumerate}
	\item Calculate $\vect{r}\si{d}$ from Eq.~\eqref{eq:rd}.
	\item If $\|\vect{r}\si{d}\|\le1$, set $\vect{r}\si{MLE}=\vect{r}\si{d}$.
	\item If $\|\vect{r}\si{d}\|>1$, find $\alpha>0$ such that $\|\vect{r}\si{MLE}(\alpha)\|=1$.
\end{enumerate}

\subsubsection{Fisher information distance}
\label{sec:chi2tomo}

When the direct inversion Bloch vector $\vect{r}\si{d}$ is only slightly outside the unit sphere of physically valid states, it may be sufficiently accurate to minimize the quadratic approximation of the Kullback--Leibler divergence~\eqref{eq:loglik},
\begin{multline}
	\label{eq:loglikapprox}
	D\si{KL}(\vect{r}\si{d}|\vect{r})
	= \frac12\left[
	\left(\frac{x-x\si{d}}{\Delta\eavg{\op{\sigma}_x}}\right)^2
	+ \left(\frac{y-y\si{d}}{\Delta\eavg{\op{\sigma}_y}}\right)^2\right.\\
	\left.+ \left(\frac{z-z\si{d}}{\Delta\eavg{\op{\sigma}_z}}\right)^2\right]
	+\mathcal{O}[(\vect{r}-\vect{r}\si{d})^3],
\end{multline}
given in terms of the error estimates of Eq.~\eqref{eq:measureerr}. This approximation, called the \emph{Fisher information distance}~\cite{GourierouxMonfort}, is easier to use than the Kullback--Leibler divergence while mostly giving comparable results (see table~\ref{tb:comparison}). In figures~\ref{fig:loglik} and~\ref{fig:loglik2} the green lines show the minima of the Fisher information distance on concentric shells around the origin $\vect{r}=0$, and the green dot in figure~\ref{fig:loglik} minimizes this distance between $\vect{r}\si{d}$ and the space of physically valid states. In analogy to Eq.~\eqref{eq:MLtrace}, the green line is given by
\begin{multline}
	\vect{r}\si{Fi}(\alpha) =\\
	 \left(\frac{x\si{d}}{1+\alpha[\Delta\eavg{\op{\sigma}_x}]^2}, \frac{y\si{d}}{1+\alpha[\Delta\eavg{\op{\sigma}_y}]^2}, \frac{z\si{d}}{1+\alpha[\Delta\eavg{\op{\sigma}_z}]^2}\right)
\end{multline}
and has similar properties, so that the three-step recipe of section~\ref{sec:likelihood} can still be used. There are situations where no $\alpha$ exists that satisfies $\|\vect{r}\si{Fi}(\alpha)\|=1$, but we have found that they are very unlikely to occur in an experiment (see tables~\ref{tb:ellipses} and~\ref{tb:comparison}).

\subsection{Maximum-likelihood estimate with radial prior}
\label{sec:MLprior}

The Bayesian prior density $\mathcal{C}(\op{\rho})$ used in Eq.~\eqref{eq:lik} contains two components that are sometimes difficult to distinguish. On the one hand, it contains a measure on the space of density matrices, which is a way of saying how ``finely grained'' this space is in its different regions, or from what distribution a purely random density matrix should be drawn in the absence of concrete knowledge about the system~\cite{Sommers2004}. This first part is likely invariant under unitary transformations (\emph{i.e.}, a Haar measure). On the other hand, $\mathcal{C}(\op{\rho})$ can contain prior knowledge about the particular situation in which we are determining density matrices, gained for example from previous experiments. This second part need not be invariant under unitary transformations. Expressed in a given parametrization, which in our case is the Bloch vector $\vect{r}$ and Eq.~\eqref{eq:rhor}, the prior density $\mathcal{C}(\vect{r})$ is the product of the measure expressed in terms of $\vect{r}$ and the density gained from prior knowledge. It is important to note that concrete prior knowledge in the absence of a measure on the space of density matrices is useless.

In the previous section we have used the Hilbert--Schmidt measure on the space of Bloch vectors~\eqref{eq:HSprior} because of its simplicity, ubiquity, and geometric appeal. However, this prejudice is misleading, and the Hilbert--Schmidt measure is neither the only nor the most natural density of quantum states of a qubit. In this section we discuss different density-matrix measures, and then use these to generalize the maximum-likelihood method to non-trivial priors.

\subsubsection{Radial prior densities of quantum states}
\label{sec:priors}

There is much freedom in defining a measure on the space of Bloch vectors. In order to focus on more natural measures, we use a physical argument for defining such a measure: a constructive procedure related to quantum state purification~\cite{Zyczkowski2001,Sommers2004,BlumeKohout2010}. 

We start from the observation that the density of \emph{pure} states of a $d$-dimensional quantum system is uniquely defined as a Haar measure over the unitary group $U(d)$; that is, since every pure state is related to every other pure state by a unitary transformation, and since all unitary transformations can be parametrized as points on the surface of a $(d^2-1)$-dimensional hypersphere, we can use the geometric measure on this hypersphere's surface as the natural measure in the space of pure states.

Next, we consider the joint tensor-product quantum state of our two-dimensional qubit ($D=2$) and a $k$-dimensional ancillary system, for a total dimension $d=D+k$. For every pure state of this $(2+k)$-dimensional system, we can trace out the ancillary dimensions to find a reduced qubit density matrix~\eqref{eq:rhor}. The reverse is also true, called quantum state purification: for every qubit density matrix~\eqref{eq:rhor} we can find a pure state of a system of $d\ge 2D=4$ dimensions, of which our state is the partial trace. This partial trace operation therefore constructs a unique measure of qubit density matrices, depending only on the ancilla dimension $k$.
Expressed as a density in the space of qubit Bloch vectors (the unit sphere), the resulting density (measure) is
\begin{equation}
	\label{eq:radialprior}
	\mathcal{C}_k(\vect{r}) =
	\begin{cases}
		\frac{\Gamma(k+\frac12)}{\pi^{3/2}\Gamma(k-1)}(1-\|\vect{r}\|^2)^{k-2} & \text{if $\|\vect{r}\|<1$}\\
		0 & \text{if $\|\vect{r}\|>1$}
	\end{cases}
\end{equation}
for $k>1$, where $\Gamma(z)$ is the Euler gamma function. As expected, this measure only depends on the length of the Bloch vector but not on its direction. The mean squared Bloch vector of this measure is $\avg{\|\vect{r}\|^2} = 3/(2k+1)$: for larger values of $k$, mixed states carry more weight than pure states.

\begin{figure}
	\begin{center}
		\includegraphics[width=0.8\columnwidth]{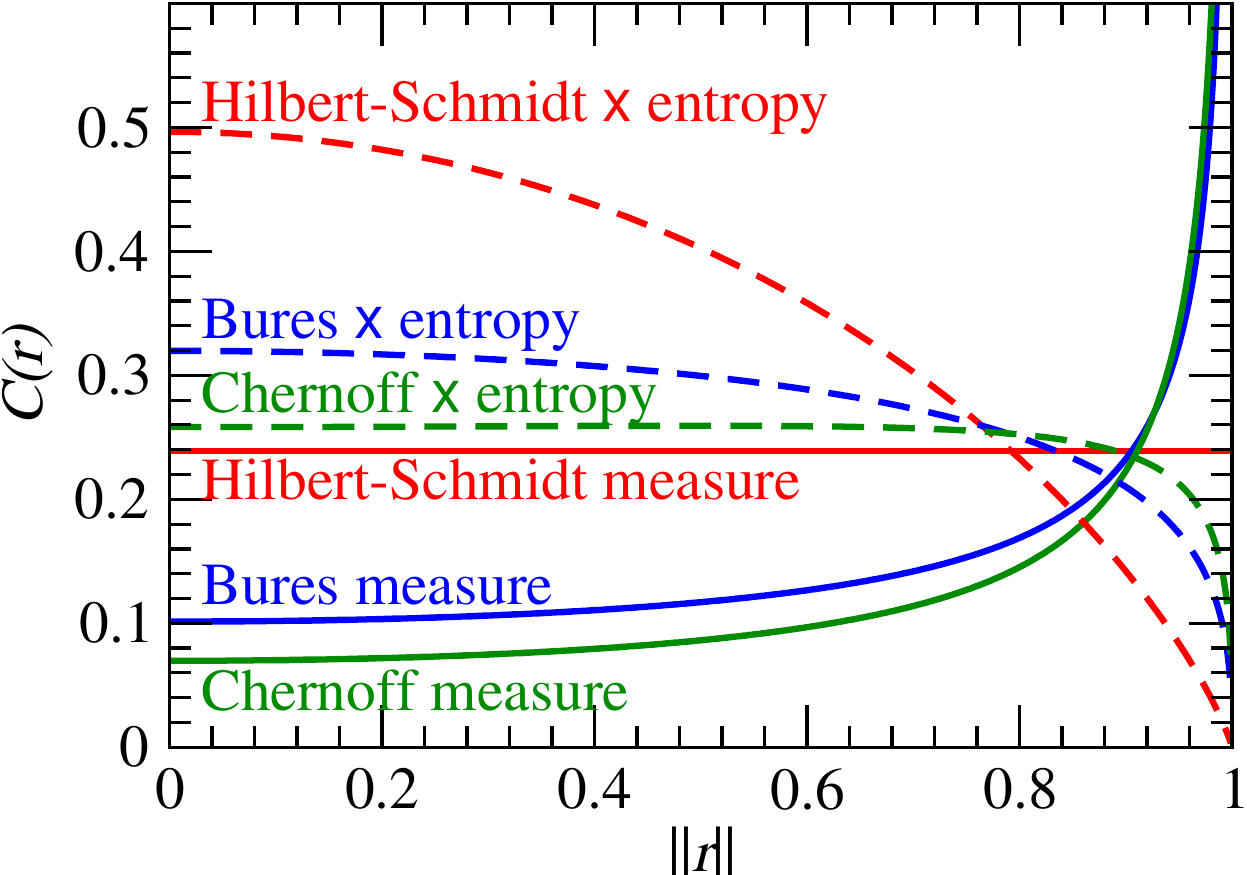}
		\caption{A few spherically symmetric prior densities of Bloch vectors. Solid red line: the Hilbert--Schmidt measure $\mathcal{C}\si{HS}(\vect{r})=\mathcal{C}_2(\vect{r})$, Eq.~\eqref{eq:HSprior}. Solid blue line: the Bures measure $\mathcal{C}\si{B}(\vect{r})=\mathcal{C}_{\frac32}(\vect{r})$, Eq.~\eqref{eq:Buresmeasure}. Solid green line: the Chernoff-information measure $\mathcal{C}\si{Ch}(\vect{r})$, Eq.~\eqref{eq:Chernoff}. The dashed lines are the same measures weighted by the entropy as in Eq.~\eqref{eq:entropyprior}, and normalized.}
		\label{fig:priors}
	\end{center}
\end{figure}

How can we choose a value for the ancillary dimension $k$? While the derivation of Eq.~\eqref{eq:radialprior} assumes that $k$ is an integer, we can use the resulting prior density for any value of $k$.
Not all values of $k$ are equally natural; we deem the following choices meaningful:
\begin{description}
	\item[$k=1$ pure states] In the limit $k\to1^+$ the measure~\eqref{eq:radialprior} becomes fully concentrated on the surface of the unit sphere ($\|\vect{r}\|=1$), meaning that only pure states have a nonzero likelihood in Eq.~\eqref{eq:lik}. While this is not a natural choice, as (strictly speaking) pure states do not exist in nature, it can be of interest for theoretical considerations or in cases where the state purity is known to be very high.
	\item[$k=\frac32$ Bures measure] In general, the Bures measure~\cite{Hall1998,Sommers2004,Osipov2010} is considered the most natural density of mixed states~\cite{Sommers2004}, as it is the Jeffreys prior~\cite{Jeffreys1946}. For qubits, its distribution is formally that of tracing over $k=\frac32$ ancillary dimensions, and has the radial density~\cite{Sentis2010}\footnote{The Bures measure for qubits~\eqref{eq:Buresmeasure} is the spherical equivalent of the Jeffreys prior of the Bernoulli trial, $\mathcal{C}(p)=1/[\pi\sqrt{p(1-p)}]$.}
	\begin{equation}
		\label{eq:Buresmeasure}
		\mathcal{C}\si{B}(\vect{r}) =
		\mathcal{C}_{\frac32}(\vect{r}) =
		\begin{cases}
			\frac{1}{\pi^2\sqrt{1-\|\vect{r}\|^2}} & \text{if $\|\vect{r}\|<1$,}\\
			0 & \text{if $\|\vect{r}\|>1$,}
		\end{cases}
	\end{equation}
	shown as a solid blue line in figure~\ref{fig:priors}.
	If nothing at all is known about the expected tomographic density matrix, then this Jeffreys prior density should be used.
	
	Note that for systems with Hilbert space dimension $D>2$, the Bures measure cannot be constructed by choosing a particular value of $k$.

	In the sphere of Bloch vectors $\vect{r}$, the Bures measure assigns a higher density of states to purer states (large $r=\|\vect{r}\|$) than to more mixed states (small $r$). We can introduce a transformed radial coordinate $s=\left[\frac{2}{\pi}\left(\sin^{-1}(r)-r\sqrt{1-r^2}\right)\right]^{1/3}$, in terms of which the Bures measure is homogeneous:
	\begin{equation}
		\label{eq:Buresmeasure_s}
		\mathcal{C}\si{B}(\vect{s}) =
		\begin{cases}
			\frac{3}{4\pi} & \text{if $\|\vect{s}\|<1$,}\\
			0 & \text{if $\|\vect{s}\|>1$.}
		\end{cases}
	\end{equation}
	This shows that the flatness of the measure depends on the chosen parametrization, and cannot be used as a criterion to prefer one measure over another.
	\item[$k=2$ Hilbert--Schmidt measure] The previously used Hilbert--Schmidt measure of Eq.~\eqref{eq:HSprior} is found by setting the ancilla dimension equal to the system dimension, $k=D=2$. It is equal to the Euclidean measure in the unit sphere of Bloch vectors, meaning that it gives every Bloch vector equal \emph{a priori} weight in the simplest geometric sense (solid red line in figure~\ref{fig:priors}). This prior is used very frequently in practice, mainly due to its mathematical simplicity; but it must be noted that it does not represent the natural density of qubit states~\cite{Sommers2004}.
	\item[$k\gg2$ highly mixed states] For large ancilla dimensions the density matrix measure becomes Gaussian and peaked around the fully mixed state,
\begin{equation}
	\mathcal{C}_k(\vect{r})\approx\left(\frac{k+\frac12}{\pi}\right)^{\frac32}e^{-(k+\frac12)\|\vect{r}\|^2}
	\text{ for $k\gg1$}
\end{equation}
	This measure can be used for tomographies where the state is known to be highly mixed.
\end{description}

There are other ways of defining a measure on the space of qubit density matrices. As an example, the Chernoff-information measure~\cite{Sentis2010}
\begin{equation}
	\label{eq:Chernoff}
	\mathcal{C}\si{Ch}(\vect{r}) = \begin{cases}
		\frac{(1-\|\vect{r}\|^2)^{-\frac12}-1}{2\pi(\pi-2)\|\vect{r}\|^2} & \text{if $\|\vect{r}\|<1$}\\
		0 & \text{if $\|\vect{r}\|>1$}
	\end{cases}
\end{equation}
follows from the experimental distinguishability of density matrices, and is shown as a green line in figure~\ref{fig:priors}.

Once a measure has been chosen for the space of density matrices, the prior density in Eq.~\eqref{eq:lik} can be taken directly from Eq.~\eqref{eq:radialprior}, \eqref{eq:Chernoff}, or other, or it can be further multiplied by a weight of our choice, for example representing concrete prior knowledge. As an example, we may use the entropy~\eqref{eq:entropy} as a radial weight, in combination with an underlying state measure:
\begin{equation}
	\label{eq:entropyprior}
	\mathcal{C}(\vect{r}) \propto \mathcal{C}_k(\vect{r}) \times S(\vect{r}),
\end{equation}
biasing the likelihood~\eqref{eq:lik} towards less pure states. Figure~\ref{fig:priors} shows a few examples of prior densities, including entropy weights.

\subsubsection{Maximum likelihood estimates with different priors}

The maximum-likelihood estimate of section~\ref{sec:likelihood} is easily adapted to any spherically symmetric prior density $\mathcal{C}(\op{\rho})=\mathcal{C}(\|\vect{r}\|)$. Since Eq.~\eqref{eq:MLtrace} maximizes the likelihood on each concentric shell $\|\vect{r}\|=\|\vect{r}\si{MLE}(\alpha)\|$, maximizing the likelihood globally thus means finding the value of $\alpha\in\mathbb{R}$ that maximizes the likelihood $\mathcal{L}[\vect{r}\si{MLE}(\alpha)]$, Eq.~\eqref{eq:lik}. For this maximization we can distinguish different classes of priors:
\begin{description}
	\item[pure or pure-peaked] If the prior density is singular at $\|\vect{r}\|=1$, for example Eq.~\eqref{eq:radialprior} with $1\le k<2$ or Eq.~\eqref{eq:Chernoff}, it is sufficient to look for the value of $\alpha$ for which $\|\vect{r}\si{MLE}(\alpha)\|=1$. Two special cases are important: if two or three components of $\vect{r}\si{d}$ are zero, then the likelihood maximum is not unique and the MLE  should not return a value; the same is true if only one component of $\vect{r}\si{d}$ is zero and the determined value of $(-\alpha)$ is larger than the number of measurements along this axis~\footnote{$J_u(t)$ has a branch cut discontinuity at $t=0$ for $u<-1$, and any results based on these values are ill-defined.}.
	\item[monotonic pure-biased] We distinguish three cases:
		\begin{description}
			\item[$\|\vect{r}\si{d}\|>1$] find the value of $\alpha>0$ for which $\|\vect{r}\si{MLE}(\alpha)\|=1$.
			\item[$\|\vect{r}\si{d}\|=1$] the likelihood maximum is at $\vect{r}\si{d}$.
			\item[$\|\vect{r}\si{d}\|<1$] find the value of $\alpha<0$, with $\|\vect{r}\si{MLE}(\alpha)\|\le1$, that maximizes the likelihood $\mathcal{L}[\vect{r}\si{MLE}(\alpha)]$.
		\end{description}
	\item[uniform (Hilbert--Schmidt)] see section~\ref{sec:likelihood} for an extended discussion.
	\item[monotonic mixed-biased] find the value $\alpha>0$ for which $\vect{r}\si{MLE}(\alpha)$ maximizes Eq.~\eqref{eq:lik}.
	\item[non-monotonic] find the value $\alpha$, with $\|\vect{r}\si{MLE}(\alpha)\|\le1$, for which $\vect{r}\si{MLE}(\alpha)$ globally maximizes Eq.~\eqref{eq:lik}.
\end{description}

In figures~\ref{fig:loglik} and~\ref{fig:loglik2} the likelihood maxima are shown for several different prior densities.
We can see that for many priors and experimental results, the MLE is rank-deficient ($\|\vect{r}\si{MLE}\|=1$), which is a serious drawback of this method~\cite{BlumeKohout2010}.

\subsection{Bayesian mean estimate}
\label{sec:BMEtomo}

Instead of reporting only the maximum of the likelihood~\eqref{eq:lik}, we can interpret the likelihood as a density in the state space and use it to calculate a weighted mean state. This Bayesian mean estimate~\cite{BlumeKohout2010} is
\begin{equation}
	\label{eq:BMEmeanrho}
	\overline{\op{\rho}}\si{BME} =
	\frac{\int \op{\rho}\,\mathcal{L}(\op{\rho})\,\mathcal{D}\!\op{\rho}}{\int \mathcal{L}(\op{\rho})\,\mathcal{D}\!\op{\rho}},
\end{equation}
where $\mathcal{D}\!\op{\rho}$ represents the chosen measure on the space of density matrices, and $\mathcal{L}(\op{\rho})$ contains the experimental knowledge including prior knowledge (see the discussion of section~\ref{sec:MLprior} on the two components of the prior density). In practice this integral is done by averaging the components of the Bloch vector,
\begin{equation}
	\label{eq:BMEmean}
	\overline{\vect{r}}\si{BME} =
		\frac{\int_{\|\vect{r}\|\le1} \vect{r}\mathcal{L}(\vect{r})\dd^3\vect{r}}{\int_{\|\vect{r}\|\le1} \mathcal{L}(\vect{r})\dd^3\vect{r}},
\end{equation}
where the measure on the space of density matrices is now \emph{included} in the definition of the likelihood~\eqref{eq:lik} and expressed in terms of the geometric Bloch vector measure $\dd^3\vect{r}$, as in Eq.~\eqref{eq:radialprior}. The Bayesian mean is generally more plausible than the likelihood maximum~\cite{BlumeKohout2010} because it is never rank-deficient.

This method can be naturally extended to higher moments of the density matrix, from which we can calculate a covariance matrix: for example, with
\begin{equation}
	\label{eq:covarianceBME}
	\overline{x^2}\si{BME} = \frac{\int_{\|\vect{r}\|\le1} x^2\mathcal{L}(\vect{r})\dd^3\vect{r}}{\int_{\|\vect{r}\|\le1} \mathcal{L}(\vect{r})\dd^3\vect{r}}
\end{equation}
we can define the variance $(\Delta x\si{BME})^2=\overline{x^2}\si{BME}-\overline{x}\si{BME}^2$, and similarly the entire covariance matrix for the components of $\overline{\vect{r}}\si{BME}$.
In figures~\ref{fig:loglik} and~\ref{fig:loglik2} we show these covariances as orange and cyan ellipses around the Bayesian mean estimates for different choices of the prior density $\mathcal{C}(\vect{r})$.

At this point we need to distinguish between two kinds of uncertainty: firstly, there is the quantum-mechanical uncertainty within a single density matrix (Bloch vector), which is typically a statistical mixture of pure states, and leads to the well-known stochastic outcomes of observables through Born's rule; and secondly, there is the uncertainty in the density matrix (Bloch vector) parametrization coming from tomographic uncertainties, given above by the covariance matrix of the components of the Bloch vector.

When we calculate the linear expectation value and variance of an operator $\op{A}$, these two types of uncertainty cannot be distinguished, and the expectation value and variance are estimated with
\begin{subequations}
\begin{align}
	\avg{\op{A}}
	&= \frac{\int_{\|\vect{r}\|\le1} \Tr[\op{\rho}(\vect{r})\op{A}]\mathcal{L}(\vect{r})\dd^3\vect{r}}{\int_{\|\vect{r}\|\le1} \mathcal{L}(\vect{r})\dd^3\vect{r}}
	= \Tr[\overline{\op{\rho}}\si{BME}\op{A}],\\
	(\Delta A)^2 &=\avg{\op{A}^2}-\avg{\op{A}}^2\nonumber\\
	&= \Tr[\overline{\op{\rho}}\si{BME}\op{A}^2]-\Tr[\overline{\op{\rho}}\si{BME}\op{A}]^2,
\end{align}
\end{subequations}
where the average density matrix is given in Eqs.~\eqref{eq:BMEmeanrho} and~\eqref{eq:BMEmean}. In this sense, the mean density matrix $\overline{\op{\rho}}\si{BME}=\op{\rho}(\overline{\vect{r}}\si{BME})$ represents the statistical mixture containing both the quantum uncertainty in each Bloch vector and the uncertainty of the parametrization of the Bloch vector itself.

The situation is different when we estimate the Bayesian mean value of a non-linear quantity such as the mean purity $\avg{\Tr(\op{\rho}^2)}$ or the mean entropy $\avg{S(\op{\rho})}$: in these cases, the covariance of the Bloch vector, Eq.~\eqref{eq:covarianceBME}, becomes important. For example,
\begin{subequations}
\begin{align}
	\avg{S(\op{\rho})} &= \frac{\int_{\|\vect{r}\|\le1} S[\op{\rho}(\vect{r})]\mathcal{L}(\vect{r})\dd^3\vect{r}}{\int_{\|\vect{r}\|\le1} \mathcal{L}(\vect{r})\dd^3\vect{r}} \neq S(\overline{\op{\rho}}\si{BME}),\\
	[\Delta S(\op{\rho})]^2 &= \frac{\int_{\|\vect{r}\|\le1} S^2[\op{\rho}(\vect{r})]\mathcal{L}(\vect{r})\dd^3\vect{r}}{\int_{\|\vect{r}\|\le1} \mathcal{L}(\vect{r})\dd^3\vect{r}}-\avg{S(\op{\rho})}^2
\end{align}
\end{subequations}
must be calculated by taking the uncertainty in the Bloch vector parametrization, given by $\mathcal{L}(\vect{r})$, into account.

\section{Error considerations}
\label{sec:errors}

\subsection{Including experimental errors}
\label{sec:smoothing}

In a real experiment, the outcomes of Stern--Gerlach measurements are never perfect. For simplicity, we assume that independently of the measurement direction, every measurement has a probability $\eta$ of giving the correct result and a probability $1-\eta$ of giving a random result (\emph{i.e.}, passing through a depolarizing channel~\cite{NielsenChuang}), or equivalently, a probability of $(1+\eta)/2$ of giving the correct result and $(1-\eta)/2$ of giving the wrong result.

Many sources of experimental errors can be expressed in this form, apart from simple detection errors. For example, if the experimental Stern--Gerlach axes fluctuate around their respective mean directions with a variance $4\avg{\sin^2(\chi/2)}$ (with $\chi$ the angle between the desired axis and the true experimental axis), the experimental error can be described by
$\eta=\avg{\cos(\chi)}=1-2\avg{\sin^2(\chi/2)}$. If several independent sources of errors $\eta_1, \eta_2,\ldots$ are present, the total error is described by their product $\eta=\eta_1\eta_2\cdots$.

In the presence of such experimental errors, the probability of measuring a certain data set is modified from Eq.~\eqref{eq:prob} to
\begin{multline}
	\label{eq:probS}
	\mathcal{P}_{\eta}(N_{x\uparrow},N_{x\downarrow},N_{y\uparrow},N_{y\downarrow},N_{z\uparrow},N_{z\downarrow} | \op{\rho}) =\\
	\binom{N_x}{N_{x\uparrow}} \left(\frac{1+\eta x}{2}\right)^{N_{x\uparrow}} \left(\frac{1-\eta x}{2}\right)^{N_{x\downarrow}}\\
	\times \binom{N_y}{N_{y\uparrow}} \left(\frac{1+\eta y}{2}\right)^{N_{y\uparrow}} \left(\frac{1-\eta y}{2}\right)^{N_{y\downarrow}}\\
	\times \binom{N_z}{N_{z\uparrow}} \left(\frac{1+\eta z}{2}\right)^{N_{z\uparrow}} \left(\frac{1-\eta z}{2}\right)^{N_{z\downarrow}}
\end{multline}
with $\eta\in[0,1]$. For $\eta=1$ the measurements are perfect and we recover Eq.~\eqref{eq:prob}; for $\eta=0$ the measurements contain no information about $\op{\rho}$.

The form of Eq.~\eqref{eq:probS} is strictly that of Eq.~\eqref{eq:prob} where the Bloch vector $\vect{r}$ is replaced by $\eta\vect{r}$. All the tomographic methods of section~\ref{sec:methods} can therefore be used to determine the vector $\eta\vect{r}$, with the caveat that the prior density depends on $\vect{r}$ and not on $\eta\vect{r}$. The direct inversion Bloch vector~\eqref{eq:rd}, which does not depend on the prior density, is now $\vect{r}\si{d}(\eta)=\vect{r}\si{d}/\eta$: it contains \emph{more} structure than $\vect{r}\si{d}$, since $\|\vect{r}\si{d}(\eta)\|\ge\|\vect{r}\si{d}\|$, in order to compensate for the loss of information during the measurement. This observation remains true for the more complicated tomographic methods discussed above, and invites the following distinction:
\begin{itemize}
	\item For $\eta<1$ we can interpret any tomographic $\vect{r}(\eta)$ as a \emph{platonic state} representing the ideal of the system, which is poorly measured in our experiment using Eq.~\eqref{eq:probS}. If we could perform a more accurate measurement, we would find a state more closely resembling this $\vect{r}(\eta)$.
	\item We can define a \emph{positivist state} $\vect{\tilde{r}}(\eta)=\eta\vect{r}(\eta)$ that already includes the effects of imprecise measurements; experimental outcomes can be predicted in terms of perfect measurements of this positivist state, using Eq.~\eqref{eq:prob}.
\end{itemize}
The author believes that platonic ideals such as $\vect{r}(\eta)$ should be discouraged in quantum mechanics, as they do not represent what can currently be measured, but instead hypothesize knowledge that may forever remain out of experimental reach. Instead, we suggest using the state $\vect{\tilde{r}}(\eta)$ as a fair representation of the experimenter's current and actual knowledge about the system.

\subsection{Estimating the tomographic uncertainty}
\label{sec:bootstrap}

\begin{figure}
	\begin{center}
		\includegraphics[width=0.8\columnwidth]{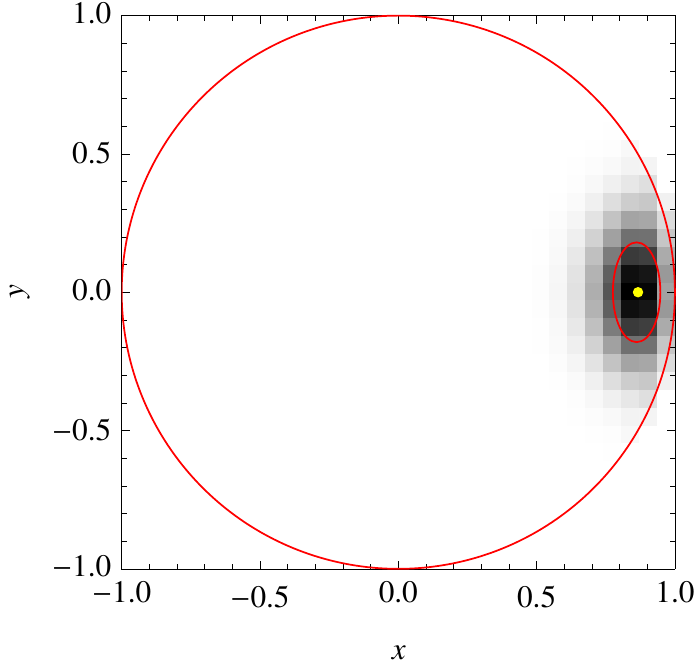}
		\caption{Bootstrapping (section~\ref{sec:bootstrap}) the state $\vect{r}=(13/15,0,0)$ (yellow dot) via scaled direct inversion tomography (section~\ref{sec:pdistance}). To generate this graphic, all 29\,791 possible experimental outcomes $(N_{x\uparrow},N_{x\downarrow},N_{y\uparrow},N_{y\downarrow},N_{z\uparrow},N_{z\downarrow})$ of performing $N_x=N_y=N_z=30$ Stern--Gerlach measurements on the state $\op{\rho}(\vect{r})$ were subjected to a tomographic state reconstruction, and the resulting Bloch vectors (projected into the $x y$ plane) were plotted in a 2D histogram using weights from Eq.~\eqref{eq:prob}, in $31\times31$ bins. The mean reconstructed state vector is at $(0.862\pm0.086,0\pm0.180,0\pm0.180)$, as listed in table~\ref{tb:ellipses}. The rms trace distance to $\vect{r}$, Eq.~\eqref{eq:accuracy}, is 0.135.}
		\label{fig:tomohist}
	\end{center}
\end{figure}

\begin{table*}
\caption{The bootstrapping covariance ellipses of figure~\ref{fig:tomohist} for the different tomography methods, using the exemplary input state $\vect{r}=(13/15,0,0)$ and $N_x=N_y=N_z=30$ measurements.
	For each method, the corresponding color in figure~\ref{fig:loglik} is indicated. 
	The ellipses are centered at $\avg{\vect{r}\si{tomo}}=\{\avg{x\si{tomo}},0,0\}$ and have the given radii $\{\Delta x,\Delta y,\Delta z\}$ in the Cartesian directions.
	$\Delta\si{tomo}$ gives the mean accuracy in terms of the rms trace distance to $\vect{r}$, Eq.~\eqref{eq:accuracy}.
	The most accurate methods (and up to 5\% higher) are highlighted in green, and poorly performing methods are highlighted in red.
	The last column gives the failure rate for methods that do not always give a well-defined result.}
\label{tb:ellipses}
\begin{tabular}{llclllll}
	\hline\hline
	\multicolumn{3}{l}{method and prior density} & $\avg{x\si{tomo}}$ & $\Delta x$ & $\Delta y,z$ & $\Delta\si{tomo}$ & $P\si{fail}$\\
	\hline\hline
	\multicolumn{2}{l}{scaled direct inversion (Sec.~\ref{sec:pdistance})} & \filleddot{1,0,0} & 0.862 & 0.086 & 0.180 & 0.135 &\\
	\hline
	\multicolumn{2}{l}{Fisher information distance (Sec.~\ref{sec:chi2tomo})} & \filleddot{0,0.667,0} & 0.866 & 0.091 & 0.168 & 0.127 & $5\times10^{-10}$\\
	\hline
	MLE: & $1\le k<2$ and Chernoff measure & \filleddot{0.5,0.5,0.5} & 0.924 & 0.045 & 0.269 & \hf 0.193 & 3\%\\
	(Sec.~\ref{sec:MLprior}) & $k=2$ (Hilbert--Schmidt) & \filleddot{0.5,0.5,0.5} & 0.864 & 0.088 & 0.174 & 0.131 &\\
	\hhline{~-------}
	& Chernoff with entropy weight & \filleddot{0,0,0} & 0.853 & 0.084 & 0.165 & 0.124 &\\
	& $k=\frac32$ with entropy weight & \filleddot{0,0,0} & 0.844 & 0.085 & 0.160 & \hc 0.122 &\\
	& $k=2$ with entropy weight & \filleddot{0,0,0} & 0.816 & 0.083 & 0.149 & \hc 0.116 &\\
	\hline
	BME: & $k=1$ (pure states) & \opendot{1,0.5,0} & 0.907 & 0.044 & 0.224 & \hf 0.161 &\\
	(Sec.~\ref{sec:BMEtomo}) & Chernoff measure & \opendot{1,0.5,0} & 0.842 & 0.101 & 0.167 & 0.129 &\\
	& $k=\frac32$ (Bures measure) & \opendot{1,0.5,0} & 0.830 & 0.077 & 0.162 & 0.122 &\\
	& $k=2$ (Hilbert--Schmidt measure) & \opendot{1,0.5,0} & 0.797 & 0.077 & 0.148 & \hc 0.117 &\\
	\hhline{~-------}
	& Chernoff with entropy weight & \opendot{0,1,1} & 0.790 & 0.084 & 0.146 & \hc 0.118 &\\
	& $k=\frac32$ with entropy weight & \opendot{0,1,1} & 0.781 & 0.076 & 0.142 & \hc 0.116 &\\
	& $k=2$ with entropy weight & \opendot{0,1,1} & 0.756 & 0.075 & 0.136 & \hc 0.117 &\\
	\hline\hline
\end{tabular}
\end{table*}

For every tomographic reconstruction, it is important to be able to give an estimate of the uncertainty of the resulting density matrix~\cite{Christandl2012}. While the Bayesian mean of section~\ref{sec:BMEtomo} gives such an estimate via Eq.~\eqref{eq:covarianceBME}, shown as orange and cyan ellipses in figures~\ref{fig:loglik} and~\ref{fig:loglik2}, the other methods presented here do not give natural error estimates.

A widely used method for nonetheless finding such an error bar, called \emph{bootstrapping}~\cite{Efron1986} or \emph{case resampling}, goes as follows: once we have tomographically determined a density matrix from experimental data, we can use this density matrix to generate new ``fake'' data sets using the same measurement operators and the probabilities of Eq.~\eqref{eq:prob} or~\eqref{eq:probS}; the argumentation is that in principle, each one of these fake data sets could have been measured, instead of the set we have measured in reality. For each such fake data set we can then do a tomography, and finally average any observables (or just the density matrix) over these tomographies. As is shown in figure~\ref{fig:tomohist} and table~\ref{tb:ellipses}, this procedure can be used to calculate the covariance matrix of the components of the tomographic Bloch vector. While these covariances correctly estimate the uncertainty we are looking for~\cite{Efron1986}, the bootstrap method contains systematic biases for the different tomographic methods, as shown in table~\ref{tb:ellipses}. Ideally, the weighted mean of all fake-data tomographies would be equal to the input state, such that we can use this technique to extract a covariance matrix without introducing a bias; however, this is not the case~\cite{Schwemmer2015}. Nevertheless, the covariance matrix found in this way can still be used in order to get an idea of the tomographic uncertainty.

We find that
the bootstrapped covariances of the Bloch vector components are slightly smaller than the covariances estimated with the Bayesian mean (section~\ref{sec:BMEtomo}) for a single experimental data set. The similar magnitudes of these two sets of error estimates lead us to the conclusion that the bootstrap method can be a valid tool for estimating tomographic uncertainties. We believe that the cautionary footnote of Ref.~\cite{BlumeKohout2010} concerning the absurd results of bootstrapping for, e.g., $(N_x,N_y,N_z)=(0,0,1)$ do not apply when several non-commuting observables are measured, as we do in this text with $N_x=N_y=N_z\ge1$.

Concerning the choice of input state $\vect{r}$ for the generation of fake data sets, the \emph{non-parametric} bootstrap method (direct re-sampling of measured data) requires us to use the direct-inversion Bloch vector $\vect{r}\si{d}$, Eq.~\eqref{eq:rd}, even if $\|\vect{r}\si{d}\|>1$ is unphysical. While this $\vect{r}\si{d}$-based non-parametric bootstrap is closest to the experimental data and may therefore be expected to be least biased, it is unrealistic in the case $\|\vect{r}\si{d}\|>1$ because it neglects the physical condition that any density matrix used for predicting experimental outcomes, including generating fake data sets, must be positive semi-definite. If we use a different Bloch vector, for example the maximum-likelihood estimate $\vect{r}\si{MLE}$, the method is called a \emph{parametric bootstrap} and is physically better justified, albeit biased.

\section{Comparison of tomography methods}
\label{sec:compare}

\begin{table*}
\caption{Comparison of the accuracies $\Delta\si{tomo}(\vect{r})$ of various tomography methods, Eq.~\eqref{eq:accuracy}. For each method, the corresponding color in figure~\ref{fig:loglik} is indicated. For a given input state (Bloch vector $\vect{r}$), $N_x=N_y=N_z=30$ measurements are simulated along each Cartesian quantization axis, and all 29\,791 possible experimental outcomes $\vect{\mathcal{N}}=(N_{x\uparrow},N_{x\downarrow},N_{y\uparrow},N_{y\downarrow},N_{z\uparrow},N_{z\downarrow})$ are fed into each tomography method (see figure~\ref{fig:tomohist} for an example), in the same way as bootstrapping (section~\ref{sec:bootstrap} and table~\ref{tb:ellipses}); finally, the root-mean-square (rms) of the trace distances $\frac12\|\vect{r}-\vect{r}\si{tomo}\|$ of the results (see section~\ref{sec:pdistance}) are computed with weights from Eq.~\eqref{eq:prob}. Smaller values indicate better accuracy of the tomography method; the most accurate methods for each input state (and up to 5\% higher) are highlighted in green, and poor methods are highlighed in red. We show exemplary results for six input states $\vect{r}$, as in table~\ref{tb:ellipses}, and the last column gives the rms of these accuracies averaged over all possible input states $\vect{r}$ with the given method's prior density using Eq.~\eqref{eq:averageaccuracy}. ($^*$) The probability for these methods to give an ill-defined result was at most $0.2\%$, except where noted; the given mean values only include well-defined results. The strongly mixed states with large failure rates only contribute minimally to $\bar{\Delta}\si{tomo}$ due to the given prior density weighting. $^{\S}$) For the six pure states along the Cartesian axes, this method always gives exactly the correct result. ($^{\dagger}$) Averages were done with the Hilbert--Schmidt measure.}
\label{tb:comparison}
\begin{tabular}{llcccccccc}
	\hline\hline
	\multicolumn{3}{l}{method and prior density} & $\vect{r}=(0,0,0)$ & $(0,0,\frac12)$ & $(0,0,0.9)$ & $(0,0,1)$ & $(1,1,0)/\sqrt{2}$ & $(1,1,1)/\sqrt{3}$ & $\bar{\Delta}\si{tomo}$\\
	\hline\hline
	\multicolumn{2}{l}{scaled direct inversion (Sec.~\ref{sec:pdistance})}				& \filleddot{1,0,0}		& 0.158			& 0.151			& 0.132					& 0.123			& \hc 0.116		& 0.114			& \bf0.137$^{\dagger}$\\
	\hline
	\multicolumn{2}{l}{Fisher information distance (Sec.~\ref{sec:chi2tomo})$^*$}		& \filleddot{0,0.667,0}	& 0.158			& 0.151			& 0.119					& \hc$0^{\S}$		& 0.126			& 0.123			& \bf0.139$^{\dagger}$\\
	\hline
	MLE: & $k=1$ (pure states)$^*$										& \filleddot{0.5,0.5,0.5}	& \hf 37\% failure	& \hf 19\% failure	& \hf 3\% failure			& \hf 2\% failure	& \hc 0.113		& \hc 0.107		& \bf0.111\\
	(Sec.~\ref{sec:MLprior}) & Chernoff measure$^*$ \eqref{eq:Chernoff}			& \filleddot{0.5,0.5,0.5}	& \hf 37\% failure	& \hf 19\% failure	& \hf 3\% failure			& \hf 2\% failure	& \hc 0.113		& \hc 0.107		& \bf0.167\\
		& $k=\frac32$ (Bures measure)$^*$ \eqref{eq:Buresmeasure}				& \filleddot{0.5,0.5,0.5}	& \hf 37\% failure	& \hf 19\% failure	& \hf 3\% failure			& \hf 2\% failure	& \hc 0.113		& \hc 0.107		& \bf0.179\\
		& $k=2$ (Hilbert--Schmidt) \eqref{eq:HSprior}							& \filleddot{0.5,0.5,0.5}	& 0.158			& 0.151			& 0.125					& \hc 0.087		& 0.117			& 0.118			& \bf0.137\\
		\hhline{~---------}
		& Chernoff with entropy weight										& \filleddot{0,0,0}		& 0.158			& 0.150			& 0.118					& \hc 0.087		& 0.118			& 0.121			& \bf0.135\\
		& $k=\frac32$ with entropy weight									& \filleddot{0,0,0}		& 0.156			& 0.148			& \hc 0.116				& \hc 0.087		& 0.120			& 0.124			& \bf0.135\\
		& $k=2$ with entropy weight										& \filleddot{0,0,0}		& 0.150			& 0.142			& \hc 0.112				& \hc 0.090		& 0.127			& 0.133			& \bf0.135\\
	\hline
	BME: & $k=1$ (pure states)											& \opendot{1,0.5,0}		& \hf 0.443		& \hf 0.306		& 0.145					& \hc 0.086		& \hc 0.111		& \hc 0.109		& \bf0.110\\
	(Sec.~\ref{sec:BMEtomo}) & Chernoff measure								& \opendot{1,0.5,0}		& \hf 0.316		& \hf 0.634		& 0.118					& \hc 0.089		& 0.124			& 0.133			& \bf{0.274}\\
		& $k=\frac32$ (Bures measure)										& \opendot{1,0.5,0}		& 0.154			& 0.149			& \hc 0.116				& \hc 0.090		& 0.121			& 0.125			& \bf0.126\\
		& $k=2$ (Hilbert--Schmidt)										& \opendot{1,0.5,0}		& 0.148			& 0.141			& \hc 0.112				& 0.095			& 0.131			& 0.136			& \bf0.131\\
		\hhline{~---------}
		& Chernoff with entropy weight										& \opendot{0,1,1}		& \hf 1.65			& \hf 1.53			& \hc 0.112				& 0.097			& 0.139			& 0.161			& \bf{1.08}\\
		& $k=\frac32$ with entropy weight									& \opendot{0,1,1}		& \hc 0.146		& \hc 0.139		& \hc 0.112				& 0.099			& 0.136			& 0.142			& \bf0.132\\
		& $k=2$ with entropy weight										& \opendot{0,1,1}		& \hc 0.141		& \hc 0.134		& \hc 0.115				& 0.106			& 0.144			& 0.151			& \bf0.133\\
	\hline\hline
\end{tabular}
\end{table*}

In this section we quantify the performance of the different tomographic methods discussed above. We use the following procedure, similar to suggestions in Refs.~\cite{HradilQSE3,BlumeKohout2010,Ng2012}, to calculate an accuracy measure for each method:
\begin{enumerate}
	\item For a given input state $\vect{r}$ and a desired number of measurements along the Cartesian axes, in our case $N_x=N_y=N_z=30$, we enumerate all $31^3=29\,791$ possible experimental outcomes $\vect{\mathcal{N}}=(N_{x\uparrow},N_{x\downarrow},N_{y\uparrow},N_{y\downarrow},N_{z\uparrow},N_{z\downarrow})$ of Stern--Gerlach measurements, together with their probabilities $\mathcal{P}(\vect{\mathcal{N}}|\vect{r})$ from Eq.~\eqref{eq:prob}.
	\item For each possible experimental outcome $\vect{\mathcal{N}}$ we reconstruct the tomographic Bloch vector $\vect{r}\si{tomo}(\vect{\mathcal{N}})$. This step is done differently for the various methods discussed in section~\ref{sec:methods}. As an example, figure~\ref{fig:tomohist} shows a 2D histogram of the Bloch vectors reconstructed  with scaled direct inversion, binned with their weight given in Eq.~\eqref{eq:prob}. Table~\ref{tb:ellipses} compares the performance of the different tomographic methods when applied to the example of figure~\ref{fig:tomohist}.
	\item For each possible experimental outcome $\vect{\mathcal{N}}$ we quantify the tomographic error through the trace distance $\frac12\|\vect{r}-\vect{r}\si{tomo}\|$ (see section~\ref{sec:Schatten}). We choose the trace distance here because it quantifies the experimental distinguishability of the two involved density matrices; but since all Schatten $p$-distances are equivalent for qubits, this is the same as the Hilbert--Schmidt distance quantifier of Ref.~\cite{HradilQSE3}.
	\item We calculate the weighted root-mean-square (rms) trace distance of all possible experimental outcomes $\vect{\mathcal{N}}$ with
\begin{equation}
	\label{eq:accuracy}
	\Delta\si{tomo}(\vect{r}) = \frac12\Bigg[\sum_{\vect{\mathcal{N}}}
		\mathcal{P}(\vect{\mathcal{N}} | \vect{r})
		\times \left\|\vect{r}\si{tomo}(\vect{\mathcal{N}})-\vect{r}\right\|^2\Bigg]^{1/2}.
\end{equation}
	In table~\ref{tb:comparison} these accuracies are shown for several tomographic methods and for several input states $\vect{r}$.
	\item These $\vect{r}$-dependent accuracies are further averaged using the prior density appropriate for each method,
\begin{equation}
	\label{eq:averageaccuracy}
	\bar{\Delta}\si{tomo} = \left[ \int_{\|\vect{r}\|\le1} \mathcal{C}(\vect{r}) \Delta^2\si{tomo}(\vect{r})\dd^3\vect{r} \right]^{1/2}.
\end{equation}
	This is a single number characterizing a given combination of a tomographic method and a prior density, without any further parameters.
	In the last column of table~\ref{tb:comparison} we show these averaged accuracies for the methods considered here, where for methods with no inherent prior we use the Hilbert--Schmidt prior $\mathcal{C}\si{HS}(\vect{r})$, Eq.~\eqref{eq:HSprior}.
\end{enumerate}
These tomographic accuracy quantifiers contain variance contributions from both the tomographic method in question and the randomness of the measurement process (quantum projection noise). However, as the latter is independent of the tomographic method, we can nonetheless use Eqs.~\eqref{eq:accuracy} and~\eqref{eq:averageaccuracy} to compare the accuracies of different tomography methods with each other. Even though the differences in the accuracies are sometimes small, we therefore compare them carefully and hope to extrapolate the findings to higher-dimensional quantum state tomographies.

We make the following observations for the different tomographic methods:
\begin{description}
	\item[Scaled direct inversion] While this method never gives the most accurate results, it is very simple, never fails, and provides a baseline against which we can compare the more complex methods. The overall performance of this method is comparable to that of the maximum likelihood estimate with Hilbert--Schmidt prior, since their results are very often the same.
	\item[Fisher information distance] This method gives good results for mixed states, but for pure states its results are worse than those of the scaled direct inversion, except on the Cartesian axes where the results are perfect. This inconsistent behavior leads us to discourage the use of this method.
	\item[Maximum likelihood estimate] The maximum likelihood method fails for priors that are singular for pure states ($k=1$, the Bures prior, and the Chernoff information prior) by not giving unique results.
	The frequently used Hilbert--Schmidt prior, on the other hand, performs well, comparable to the scaled direct inversion. When mixed states are known to predominate, adding an entropy weight gives even slightly more accurate results.
	
	It is often argued that the maximum likelihood method with Hilbert--Schmidt prior (section~\ref{sec:likelihood}) is the ``best'' method since we cannot gain by giving an answer that is less likely, such as we do when giving a Bayesian mean estimate~\cite{HradilQSE3}. This argument is misleading, however. Firstly, the question of which prior density $\mathcal{C}(\op{\rho})$ to use in the definition of the likelihood~\eqref{eq:lik} is not answered to our satisfaction by tacitly using the Hilbert--Schmidt prior~\eqref{eq:HSprior}, especially in situations where the experimenter knows \emph{a priori} that the generated states are nearly pure. The argument that the flatness of the Hilbert--Schmidt prior makes it most natural is contingent on the chosen parametrization, see Eq.~\eqref{eq:Buresmeasure_s} and Ref.~\cite{Shang2013}. While the more natural Bures prior~\eqref{eq:Buresmeasure} fails to give satisfactory results, see table~\ref{tb:comparison}, other priors are possible, and this degree of freedom casts at least some uncertainty on the optimality of the MLE method with HS prior. Secondly, as discussed below, other methods give \emph{on average} more accurate tomographic results according to our quantifiers~\eqref{eq:accuracy} and~\eqref{eq:averageaccuracy}.
	\item[Bayesian mean estimate] The BME is not only more plausible than the MLE  because it is of full rank~\cite{BlumeKohout2010}, but according to our overall quantifier~\eqref{eq:averageaccuracy} the BME is the most accurate method studied here (except when using the Chernoff information measure, see below).
	As we can see in figures~\ref{fig:loglik} and~\ref{fig:loglik2}, the BME is strongly influenced by the choice of the prior density $\mathcal{C}(\op{\rho})$: in general, the MLE is not even contained within the corresponding BME uncertainty ellipsoid.
	In table~\ref{tb:comparison} we see that in experiments where pure states are expected, using  a pure-state ($k=1$) prior gives the best results of our study; if the purity of the state is not known \emph{a priori}, using a Bures or Hilbert--Schmidt prior is the optimal choice.
\end{description}

It may be surprising that we find the Bayesian mean estimate to be more accurate \emph{on average} than the likelihood maximum, even though the former suffers from rather strong prior-dependent biases and the latter has been shown to be the most efficient estimation strategy~\cite{HradilQSE3}. This discrepancy comes from the observation, seen in table~\ref{tb:comparison}, that while the BME is generally more accurate for mixed states, the MLE method is more accurate for pure states; however, in our averaging procedure, Eq.~\eqref{eq:averageaccuracy}, mixed states carry much more weight than the pure states near the surface of the sphere of Bloch vectors. In real experiments, the experimenter often tries to generate rather pure quantum states, and for these the MLE method indeed does give more accurate results if the Hilbert--Schmidt prior is assumed. However, we argue that in this case the Hilbert--Schmidt prior is not the correct one to use, but either the pure-state prior $\mathcal{C}_1(\vect{r})$ or the Bures prior $\mathcal{C}\si{B}(\vect{r})$, which correctly prioritize pure states; and in these cases, the BME \emph{does} out-perform the MLE method~\cite{BlumeKohout2010}, quantitatively for $\mathcal{C}_1(\vect{r})$ and qualitatively for $\mathcal{C}\si{B}(\vect{r})$ (since in this case the MLE method always returns a pure state, which is not justified \emph{a priori}).

We make the following observations for the different prior densities:
\begin{description}
	\item[$k=1$ pure states] If we can be sure \emph{a priori} that the experimental data has been generated by measurements on a pure state, then the BME method with prior $\mathcal{C}_1(\op{\rho})$ is slightly more accurate on average than the MLE; also, the MLE method has a small probability of not giving a unique result at all. Therefore, the BME is preferred in this case, keeping in mind that the BME is never a pure state.
	\item[$k=\frac32$ Bures measure] On average, the BME method gives much more accurate results than the MLE method, and it never fails. For pure states, however, the MLE is more accurate than the BME; but in this case the pure-state prior is more appropriate.
	\item[$k=2$ Hilbert--Schmidt measure] On average, the BME method gives slightly more accurate results than the MLE method. Again, for pure states the MLE is more accurate than the BME; but in this case the Hilbert--Schmidt measure is an inappropriate choice.
	\item[Chernoff information measure] While the Chernoff information measure gives good results for the MLE, comparable to those of the Bures measure, it gives very poor results for the BME of mixed states. The reason for this is that the Chernoff information measure is strongly peaked at pure states; but even for the BME of pure states its results are less accurate than those obtained with the Bures measure. For this reason we discourage the use of the Chernoff information measure.
	\item[Entropy weights] In general, both the MLE and the BME give very good results with entropy-weighted prior densities. An exception is the entropy-weighted Chernoff information measure, which gives very poor results for mixed states.
\end{description}

We conclude that the Bayesian mean estimate is the preferred method for single-qubit quantum state tomography. The instances where the maximum-likelihood method performs better, namely when pure states are reconstructed with the Hilbert--Schmidt or Bures prior, are not well justified since these priors are ill adapted to the experimental situation concerning pure states.

\subsubsection{Extrapolation to larger systems}
\label{sec:largersystems}

For systems with larger Hilbert spaces (dimension $D\gg2$), the direct inversion result is very likely to be non-physical. The reason for this is that experimental quantum states are mostly of very low rank, and the tomographic estimates of the zero eigenvalues of the density matrix are statistically scattered around zero~\cite{Wigner1958}, with the probability of all of them being positive becoming exponentially small with the system dimension.

For this reason, the MLE becomes independent of the choice of prior, since any measure $\mathcal{C}_k$ for $k\le D$, as well as the Bures measure, will yield the same rank-deficient result (the equivalent of a Bloch vector on the surface of the unit sphere for the single-qubit case, the gray dot in figure~\ref{fig:loglik}). In this sense, the usual choice of the Hilbert--Schmidt measure $k=D$~\cite{HradilQSE3} is valid.

For $D\gg2$, the BME becomes computationally difficult to evaluate and must be calculated with a Monte Carlo algorithm. While in principle the BME is still the preferred method~\cite{BlumeKohout2010}, such practical difficulties may discourage its use in large systems.

In our experimental practice with two-component Bose--Einstein condensates~\cite{RiedelNATURE2010,Schmied2011b,OckeloenPRL2013} we apply these insights and use the MLE with Hilbert--Schmidt prior for quantum state reconstruction, as is done in other groups~\cite{Strobel2014}. We have found that this is the only computationally feasible and physically valid method for systems comprising hundreds or even thousands of particles, and consider it fortuitous that the present comparative study deems it appropriate. The problem of its general rank-deficiency is considered acceptable.

\begin{acknowledgments}
The author would like to thank Philipp Treutlein, Andreas Nunnenkamp, and Christoph Bruder for valuable discussions and criticism. The Centro de Ciencias de Benasque Pedro Pascual has provided a very stimulating environment for finishing this manuscript. This work was supported by the Swiss National Science Foundation and by the EU project QIBEC.
\end{acknowledgments}

\appendix

\section{Choice of measurement axes}
\label{app:uniform}

In this section we motivate the set of Cartesian measurement axes used for spin-$1/2$ tomography in this work (see section~\ref{sec:intro}). We find the conditions under which a set of measurement axes can be used for efficient and accurate quantum state tomography.

\subsection{Direct inversion (filtered backprojection)}
\label{app:fbp}

The direct inversion (filtered backprojection) technique of Ref.~\cite{Schmied2011b} is the simplest general tomographic method for quantum-mechanical spins of arbitrary length $j$. Given a true density matrix with spherical tensor coefficients $\rho_{k q}$ and a set of $M$ measurement axes $(\vartheta_n,\varphi_n)$ with measurement weights $c_n$, the average tomographically reconstructed spherical coefficients of the density matrix (averaged over all possible measurement results) are found by inserting Eq.~(6) into Eq.~(4) of Ref.~\cite{Schmied2011b},
\begin{multline}
	\avg{\rho_{k q}\se{(fbp)}}\\
	= (2k+1)\sum_{n=1}^M c_n D_{q 0}^k(\varphi_n,\vartheta_n,0)\sum_{m=-j}^j p_m(\vartheta_n,\varphi_n)t_{k 0}^{j m m}\\
	= \sum_{q'=-k}^{k} \rho_{k q'}
	\times 
	4\pi\sum_{n=1}^M c_n
	\left[Y_k^q(\vartheta_n,\varphi_n)\right]^*
	Y_k^{q'}(\vartheta_n,\varphi_n).
\end{multline}
This is the correct result $\avg{\rho_{k q}\se{(fbp)}}=\rho_{k q}$ if the measurement axis orientations $(\vartheta_n,\varphi_n)$ and their weights $c_n$ satisfy
\begin{equation}
	\label{eq:fbpcorrect}
	4\pi\sum_{n=1}^M c_n \left[Y_k^q(\vartheta_n,\varphi_n)\right]^* Y_k^{q'}(\vartheta_n,\varphi_n) = \delta_{q q'}
\end{equation}
for all $k=0,1,\ldots,2j$ and $q,q'=-k,-k+1,\ldots,+k$. If we decompose the angular density of measurement axes into a sum of spherical harmonics with coefficients
\begin{equation}
	s_{k q}
	= \sum_{n=1}^M c_n \left[Y_{k}^{q}(\vartheta_n,\varphi_n)\right]^*,
\end{equation}
then Eq.~\eqref{eq:fbpcorrect} is satisfied whenever $s_{k q}=0$ for all $k=2,4,6,\ldots,4j$.

For $j=1/2$, as used in this work, this implies that direct inversion tomography~\cite{Schmied2011b} is correct on average if the distribution of measurement axes satisfies $s_{2,-2}=s_{2,-1}=s_{2,0}=s_{2,1}=s_{2,2}=0$, \emph{i.e.}, if there is no quadrupolar anisotropy in the distribution of measurement axes. The smallest set of measurement axes that satisfies these conditions is the Cartesian-axes tomography set with equal numbers of measurements along each axis, as used in this text.

\subsection{Maximum likelihood estimation}

In contrast to direct inversion (section~\ref{app:fbp}), maximum likelihood estimation is intrinsically biased due to the physicality constraint on the density matrix (see table~\ref{tb:ellipses}). We can nonetheless ask: what conditions must a set of measurement axes fulfil so that the mean tomographic estimate is closest to the true density matrix?

For the tomographic reconstruction of a completely mixed spin-1/2 state (Bloch vector $\vect{r}=0$), we can quantify this question by calculating the variance $\avg{\|\vect{r}\si{MLE}\|^2}$ of the resulting Bloch vector, averaged over all possible sets of experimental results. We find numerically that $\avg{\|\vect{r}\si{MLE}\|^2}$ is smallest if the distribution of measurement axes has $s_{2,-2}=s_{2,-1}=s_{2,0}=s_{2,1}=s_{2,2}=0$, which is the same vanishing-quadrupole condition as found in section~\ref{app:fbp}.

\bibliography{spinhalftomo}

\end{document}